\documentclass[aps,prb,twocolumn,amsmath,showpacs]{revtex4}
\usepackage{graphicx}

\newcommand{\EqLabel}[1] { \label{#1} }

\begin{document}

\title{Magnetic susceptibility of diluted magnetic semiconductors at
  low carrier densities}

\author{Adel Kassaian}

\author{Mona Berciu}

\affiliation{Department of Physics and Astronomy, University of
British Columbia, Vancouver, BC V6T 1Z1, Canada }

\date{\today}

\begin{abstract}
We calculate the static longitudinal and the transverse dynamic
magnetic susceptibilities of (III,Mn)V diluted magnetic
semiconductors, using the Random Phase Approximation, for a simple
impurity band model appropriate for the low charge carrier
concentration regime. The magnetic susceptibilities are shown to
depend sensitively on the amount of positional disorder of the Mn
impurities. The results we obtain are consistent with previous studies
of the spin wave spectrum and of the spatially inhomogeneous
ferromagnetic state of these materials.
\end{abstract}

\maketitle

\section{Introduction} 

Diluted magnetic semiconductors (DMSs) are obtained by doping a
semiconductor with magnetic impurities. To date, Ga$_{1-x}$Mn$_x$As
has been the most studied\cite{Ohnorev} III-V DMS because it has the
highest reliable critical temperatures recorded: 160K for bulk
samples\cite{Edmonds} and 172K in digitally doped
heterostructures.\cite{Nazmul} In Ga$_{1-x}$Mn$_x$As, substitution of
a fraction $x$ of the Ga with Mn introduces both local Mn spins
($S=5/2$) and holes into the system.  It is widely accepted that
magnetization is due to charge-carrier mediated, effectively
ferromagnetic (FM), interactions between the Mn
spins.\cite{Ohnorev,Beschoten} It is known that these alloys are
heavily compensated, with a hole concentration much smaller than the
Mn concentration.  DMSs are alloys, with inherent positional disorder
of Mn atoms. Other types of defects, such as As antisites and Mn
interstitials, are also present.\cite{Ohnorev,Yu} The spin-orbit
coupling may play a significant role by making these interactions
anisotropic,\cite{Janko,Gerg} although it is not clear to what
extent.\cite{Brey,Zhou} A theoretical treatment which fully takes into
account all these factors is not yet available. Instead, theoretical
work tends to focus on different aspects of the problem.

Our recent work\cite{MB1,HF,RPA,MB2,MPK1,MPK2} has been focused on
understanding the effects of {\em positional disorder} of the Mn
impurities on the magnetic properties of these compounds. Disorder is
known to induce localization of the states lying at the bottom of the
band, below the mobility edge. When the Fermi energy crosses the
mobility edge, the system undergoes a metal-insulator transition
(MIT), at $x\sim 0.03$ in GaMnAs.\cite{Ohnorev} Since transport
properties (metal vs. insulator) are determined by the nature of the
states near the Fermi energy (extended vs. localized) one might argue
that for the samples with the highest $T_c$, which are also the most
metallic ones,\cite{Edmonds} disorder effects are
unimportant. However, the main applications of these materials are
based on their magnetic properties. Unlike transport properties, the
magnetic properties depend on the nature of {\em all} the occupied
states, not only the ones near the Fermi energy, since {\em all} the
charge carriers interact with the Mn spins. Disorder may thus
influence magnetic properties considerably, certainly on the
insulating side, but also above the MIT (even the most metallic GaMnAs
samples have very short mean free paths).

In this work, we investigate the effects of positional disorder on the
magnetic susceptibility of these materials.  The model we
study\cite{MB1,HF,RPA,MB2,MPK2} is an impurity band model.  It is
expected to be (at least qualitatively) valid at low concentrations,
below and near the MIT, where the disorder effects are likely to be
largest and thus most easy to identify. Although we use GaMnAs as a
prototype, other insulators, such as GaMnN and GeMn may exhibit
similar physics, if they are indeed DMS.\cite{Heb,GeSi} In order to
understand the effects of positional disorder, we contrast the
behavior of ordered samples (where the Mn are assumed to be placed on
an ordered cubic superlattice) with weakly, moderate and fully
disordered configurations, where we allow the randomness in the Mn
positions to increase gradually.\cite{MB1} It should be emphasized
that the results for the ordered systems also apply to itinerant
models,\cite{Konig,Amir,Dietl} provided that the appropriate mapping
of parameters (discussed below) is performed. The method we employ is
the Random Phase Approximation (RPA); this, and the low density regime
we consider distinguish our work from other recent computations of
magnetic susceptibilities, based on Boltzmann
equations.\cite{Timmetal}

The paper is organized as follows: in Section II we briefly review the
model and the self-consistent mean-field solution. In Section III we
discuss the static longitudinal susceptibility. In Section IV we
derive the generalized Random Phase Approximation equations, which are
used in section V to compute the dynamical transverse susceptibility
for ordered and disordered systems. Finally, Section VI contains our
conclusions.

\section{The Model and the Mean-Field Approximation}

The model we investigate has been described in detail in
Refs. \onlinecite{MB1,HF,MPK2,RPA}. We briefly review it here. The
host is assumed to have zinc-blende structure.  $N_d$ Mn dopants are
placed at positions $\vec{R}_i$, $i=1,...,N_d$ on an $N\!\times\! N
\!\times\! N$ FCC sublattice, of lattice constant $a$ ($=5.65~\AA$ for
GaAs), corresponding to a doping $x = N_d/4N^3$.  The number of charge
carriers is fixed to $N_h=pN_d$, where $p <1$ due to compensation. We
use periodic boundary conditions.  The Hamiltonian we investigate is:
\begin{equation}
\label{1.1}
{\cal H}(t)\! = \!\!\sum_{i,j,\sigma}^{}t_{ij} c^{\dagger}_{i\sigma}
c_{j\sigma} + \!\sum_{i,j}^{}J_{ij} \vec{S}_i \cdot \vec{s}_j -g
\mu_{B} \!\!\sum_{i} \vec {B}(i,t)\cdot(\vec{s}_i+\vec{S}_i)
\end{equation}
Here, $c^{\dagger}_{i\sigma}$ creates a charge carrier with spin
$\sigma$ in the impurity state centered at $\vec{R}_i$. The first term
describes hopping of charge carriers between impurity states, where
$t_{ij} = 2(1+r/a_{\rm B})\exp{(-r/a_{\rm B})}$ Ry, with $r=|{\bf R}_i
- {\bf R}_j|$.\cite{Bhatt1} For Mn in GaAs, 1 Ry$\sim110$ meV and
$a_B\approx 8\AA$. \cite{BG,MB1} This particular hopping Hamiltonian
has been shown to describe an impurity band which has a mobility edge,
as well as a characteristic energy for the occupied states in
agreement with physical expectations (a detailed discussion of these
issues is presented in the Appendix of Ref.~\onlinecite{HF}, as well
as Ref.~\onlinecite{MB2}). The second term describes the
antiferromagnetic exchange between the Mn spin ${\vec S}_i$ and the
charge carrier spin ${\vec s}_j={ 1\over 2} c^{\dagger}_{j\alpha}
{\vec \sigma}_{\alpha\beta} c_{j\beta}$ (${\vec \sigma}$ are the Pauli
spin matrices), which is proportional to the probability of overlap
between the charge carrier trapped at ${\vec R}_j$ and the Mn spin at
${\vec R}_i$: $J_{ij} = J \exp{(-2 |{\vec R}_i-{\vec R}_j|/a_B
)}$. The exchange between a hole and its own Mn
($\vec{R}_i=\vec{R}_j$) is $J=15$ meV.\cite{MB1,BG} The third term
describes the coupling to an external magnetic field.  For simplicity,
we assume that both types of spins have the same $g$-factor. The value
of the holes' $g$ is unimportant, because the magnetic properties are
dominated by the Mn spins.\cite{HF}

This Hamiltonian obviously neglects several other possible
terms. Since the system is heavily compensated, it must contain a
significant amount of charged compensation centers. The electric
potential created by these charged defects leads to the appearance of
a disordered on-site energy, of type $\sum_{i\sigma}^{} \epsilon_i
c^{\dagger}_{i\sigma} c_{i\sigma}$. The spread in the distribution of
$\epsilon_i$ is dependent on the amount of correlations between the
positions of these charged defects, established during
growth.~\cite{Timm} A Hubbard-like term $U\sum_{i}^{}n_{i\uparrow}
n_{i\downarrow}$ should be added to limit double-occupancy of the
impurity states. Since the charge carrier density is low, one could
argue that in fact longer range electron-electron repulsions are
needed. We have studied the effect of adding such terms, as well as
modeling differently the hopping term, in Ref.~\onlinecite{HF}. They
are found to lead to some quantitative changes, but no qualitatively
new physics.  As we propose to focus on the effects of positional
disorder on the magnetic susceptibilities, we ignore such extra terms
here.  In Hamiltonian (\ref{1.1}) we also assume that the impurity
states have the simple s-wave symmetry typical of donor levels,
ignoring the more complicated structure of impurity acceptor levels
due to the multi-band valence band-structure.~\cite{BG} Unless the
spin-orbit coupling is very strong, we believe that this approximation
also leads to only quantitative changes. The formalism we develop here
can be straightforwardly generalized to take all these extra terms
into account; however, we do not expect qualitative changes to the
results we report.

We first consider a homogeneous, static external magnetic field, $
\vec {B}(i,t)=B \vec{e}_z$. The mean-field solution, based on the
customary factorization of the interaction term, was investigated in
Refs. \onlinecite{MB1,HF}.  We rederive it here using a variational
approach. Then, we generalize this approach to spatial/time-dependent
external fields, to find the RPA equations and dynamic response
functions.

The idea is to replace the full interacting Hamiltonian
\begin{equation}
\EqLabel{new1} {\cal H}^{B}\! = \!\!\sum_{i,j,\sigma}^{}t_{ij}
c^{\dagger}_{i\sigma} c_{j\sigma} + \!\sum_{i,j}^{}J_{ij} \vec{S}_i
\cdot \vec{s}_j -g \mu_{B} B\!\!\sum_{i} ({s}^z_i+{S}^z_i)
\end{equation}
with the particular quadratic form\cite{note1}
\begin{equation}
\label{12}
\hat{\cal K}^B = \sum_{ij,\sigma}^{}h^{B}_{ij,\sigma}
c^{\dagger}_{i\sigma} c_{j\sigma} - \sum_{i}^{}H^{B}_{i} S_i^z
\end{equation}
which minimizes the free energy\cite{Ripka} ${\cal F}(\hat{\cal
K}^B)\ge {\cal F}_{eq} $, where
\begin{equation}
\label{13}
{\cal F}(\hat{\cal K}^B) = -k_{B} T \ln {\cal Z}_{0}^B +Tr \{\hat{\cal
D}_{0}^B [\hat{\cal H}_0^B-\hat{\cal K}^B]\}.
\end{equation}
Here, $\hat{\cal N}=\sum_{i, \sigma} c^{\dagger}_{i\sigma}
c_{i\sigma}$ is the particle number operator, $ \hat{\cal
D}_{0}^B=\exp[-\beta (\hat{\cal K}^B- \mu^B \hat{\cal N})]/{\cal
Z}_{0}^B$ is the trial density matrix, where ${\cal Z}_{0}^B= Tr
\{\exp{[-\beta (\hat{\cal K}^B-\mu^{B} \hat{\cal N})]}\}$, and $\mu^B$
is the chemical potential. We use the upper index $B$ to distinguish
between solutions in different static external magnetic fields
$B\vec{e}_z$. If $B=0$ we will drop this index.

We define the expectation values:
\begin{eqnarray}
\label{16}
&\rho_{ji,\sigma}^{B}= Tr \{\hat{\cal D}_{0}^B c^{\dagger}_{i\sigma}
c_{j\sigma}\} = -\frac{1}{\beta}^{} {\partial \ln {\cal
Z}_0^B}/{\partial h^B_{ij,\sigma}} \\
&\label{17}
\langle\vec{S}_{i}\rangle=Tr\{\hat{\cal D}_{0}^B \vec{S}_{i}\}= S_{\rm
Mn}^B(i)\hat{e}_{z}
\end{eqnarray}
where
\begin{equation}
\label{17a}
{S^B_{\rm Mn}(i)}= \frac{1}{\beta}^{} \frac{\partial \ln {\cal
Z}_{0}^B}{\partial H_{i}^B}=B_{S}(\beta H_i^B).
\end{equation}
and $B_{S}(x)=(S+{1 \over 2}) \coth[(S+{1 \over 2})x]-{1 \over 2}
\coth{x \over 2} $ is the Brillouin function ($S={5 \over2}$ for
Mn). We then have:
\begin{eqnarray*}
 &{\cal F}(\hat{\cal K}^B) = -k_{B} T \ln {\cal
 Z}^B_{0}+\sum_{ij,\sigma} t_{ij} \rho^B_{ji,\sigma} \\
 +\sum_{ij,\sigma} J_{ij}& S^B_{\rm Mn}(i) {\sigma \over 2}
 \rho^B_{jj,\sigma} - g \mu_B B \sum_{i}^{}[S^B_{\rm
 Mn}(i)+\sum_{\sigma}^{}{\sigma \over 2} \rho^B_{ii,\sigma}]\\
 &-\sum_{ij,\sigma} h^B_{ij,\sigma} \rho^B_{ji,\sigma} +\sum_{i}
 H^B_{i} S^B_{\rm Mn}(i).
\end{eqnarray*}
The variational parameters $h^{B}_{ij,\sigma}$ and $H^{B}_{i}$ are
obtained straightforwardly\cite{Ripka} from the minimization $\delta
{\cal F}=0$:
\begin{eqnarray}
\label{mf1}
 &H^B_{i}= g \mu_B B -\sum_{j,\sigma} \frac{\sigma}{2} J_{ij}
 \rho^B_{jj, \sigma} \\
\label{mf2} &h^B_{ij,\sigma}= t_{ij}+\frac{\sigma}{2} \delta_{ij} \left[\sum_{k}J_{ik}
S^B_{\rm Mn}(k)- g \mu_B B\right]
\end{eqnarray}
These are the self-consistent mean-field equations. They can be
written in the familiar form if the electronic part of the trial (or
mean-field) Hamiltonian is diagonalized:
\begin{equation}
\label{24}
{\cal K}^B_{el}=\sum_{ij,\sigma} h^B_{ij,\sigma} c^{\dagger}_{i\sigma}
c_{j\sigma}=\sum_{n\sigma}^{} E^B_{n\sigma} a^{\dagger}_{n\sigma}
a_{n\sigma}
\end{equation}
through a unitary transformation:
\begin{equation}
\label{22}
a^{\dagger}_{n\sigma} = \sum_{i}^{} \psi^B_{n\sigma}(i)
c^{\dagger}_{i\sigma}.
\end{equation}
The diagonalization condition is:
\begin{equation}
\label{25}
\sum_{j} h^B_{ij,\sigma}\psi^B_{n\sigma}(j)=
E^B_{n\sigma}\psi^B_{n\sigma}(i).
\end{equation}

These equations determine the self-consistent solution. We start with
an initial guess for the values of $S^B_{\rm Mn}(i)$ (see
Ref. \onlinecite{HF} for details). We use Eqs.  (\ref{mf2}) and
(\ref{25}) to find the fermionic energies $E^B_{n\sigma}$ and wave
functions $\psi^B_{n\sigma}(i)$. Using Eqs. (\ref{16}), (\ref{24}) and
(\ref{22}), the fermionic fields become:
\begin{equation}
\label{26a}
\rho^B_{jj,\sigma}=\sum_{n} f(E^B_{n \sigma})|\psi^B_{n\sigma}(j)|^2
\end{equation}
where $ {f(E)} = [\exp(\beta(E-\mu^B))+1]^{-1}$ is the Fermi
distribution and the chemical potential $\mu^B$ is given by:
\begin{equation}
\label{27}
\sum_{i,\sigma}^{}\rho^B_{ii,\sigma}=\sum_{n,\sigma}f(E^B_{n\sigma})=N_{h}
\end{equation}
Once the fermionic fields are known, using Eqs. (\ref{17a}) and
(\ref{mf1}) we can obtain the new spin expectation values $S^B_{\rm
Mn}(i)$. We repeat the iterations until self-consistency is reached.

\section{The static longitudinal susceptibility}

This response function characterizes the change in the total
magnetization, when a static external magnetic field is applied
parallel to the magnetization axis. We can separate it into two
components, $\chi = \chi_{\rm Mn} + \chi_h$, where
$$ \chi_{\rm Mn} = {g\mu_B \over N_d}\sum_{i}^{}\left.{d S^{B}_{\rm
Mn}(i) \over d B}\right|_{B=0}; \chi_h= {g\mu_B \over
N_d}\sum_{i}^{}\left.{d s^{B}_{h}(i) \over d B}\right|_{B=0}
$$ (To obtain the susceptibility per unit volume, one needs to
multiply by $n_{\rm Mn}=4x/a^3$). The charge carrier spin expectation
values are [Eqs. (\ref{16}), (\ref{26a})] $s^{B}_{h}(i)=
\sum_{\sigma}^{} \sigma \rho^B_{ii,\sigma}/2$.

One method to calculate these susceptibilities is by direct numerical
evaluation of this derivative, {\em e.g.}
\begin{eqnarray}
\label{1.20}
\chi_{\rm Mn} \approx g \mu_B\left.{{S^{B}_{\rm Mn}-S_{\rm Mn}} \over
  B} \right|_{g \mu_B B/J \ll 1}
\end{eqnarray}
where $S^{B}_{\rm Mn}=\sum_{i}^{}S^{B}_{\rm Mn}(i)/N_d$ and $S_{\rm
Mn}=S^{B=0}_{\rm Mn}$ are the average Mn spin with and without a
static magnetic field, which can be obtained directly from the
mean-field solutions. The main issue with this approach is the proper
choice of $B$ and the proper self-consistency criterion to be
used. Clearly, $S_{\rm Mn}$ and $S^{B}_{\rm Mn}$ must be computed to
very high accuracy so that errors in the numerator of Eq. (\ref{1.20})
are small relative to the small value of $B$ chosen. We obtain good
convergence with results of another method (described below) for $g
\mu_BB=10^{-4}$ meV and self-consistency defined by the condition that
the variation of the total magnetization in successive iterations is
less than $10^{-6}$. While reaching such high accuracy is time
consuming, this method is the most efficient way to compute the static
susceptibility for large system ($N_d > 500$).

The more customary way to compute a linear response function, however,
is to express it in terms of expectation values of the unperturbed
system (i.e., $B=0$ quantities). Let us first derive the longitudinal
susceptibility for an ordered system, i.e. one where the Mn impurities
are assumed to be placed on a simple cubic superlattice inside the
host semiconductor. In this case, due to invariance to translations,
we have $S^B_{\rm Mn}(i) = S^B_{\rm Mn}$, $s^B_h(i)=s_h^B, \forall
i$. The charge carrier part of the mean-field Hamiltonian
[Eqs. (\ref{mf2}), (\ref{24})] has eigenfunctions which are
plane-waves for all $\vec{k}$ inside the first Brillouin zone. We
find:
\begin{eqnarray}
\label{4.66}
E^{B}_{\vec{k}\sigma} &=& \epsilon_{\vec{k}} + {\sigma \over 2} ({ J_0
S^{B}_{\rm Mn}- g \mu_{B} B}) \\
\label{5.66}
 s^{B}_{h} &=& {1\over 2 N_d}\sum_{\vec{k} \sigma} \sigma
f(E^{B}_{\vec{k} \sigma}) \\
\label{6.66}
S^{B}_{\rm Mn} &=& B_S\left[\beta({g} \mu_{B} B - J_{0}
s^{B}_{h})\right].
\end{eqnarray}
Here, $\epsilon_{\vec{k}} = \sum_{\vec{\delta}\ne 0}^{}
t_{\vec{\delta}} \exp{(i\vec{k}\cdot\vec{\delta})}$ is the kinetic
energy of the non-interacting electrons, where $t_{\vec{\delta}}=
t_{ij}$ for which $\vec{R}_i - \vec{R}_j = \vec{\delta}$. Also, $J_{0}
= \sum_{\vec{\delta}} J_{\vec{\delta}}$, where
$J_{\vec{\delta}}=J_{ij}$.

From Eq. (\ref{6.66}), we find the spin contribution:
\begin{equation}
\label{2.10}
{\chi_{\rm Mn}\over g\mu_B}=\beta \left(g\mu_B - J_{0} {\chi_{h} \over
g\mu_B}\right) {{B'_S}}(-\beta J_{0} s_{h})
\end{equation}
where $B'_S(x) = {d \over dx} B_S(x)$. From Eq. (\ref{5.66}), we have:
\begin{equation}
\label{5.10}
\chi_{h} = {g \mu_B \over 2 N_d} \sum_{\vec{k} \sigma} \sigma
\left.\left({d E^{B}_{\vec{k}\sigma} \over d B}-{d \mu^B \over d B}
\right)\right|_{B=0} g(E_{\vec{k} \sigma})
\end{equation}
where $g(E)={d \over dE} f(E)$. From Eq. (\ref{4.66}) we find:
\begin{equation}
\label{3.10}
\left.{d E^{B}_{\vec{k}\sigma} \over d
  B}\right|_{B=0}=E^{(1)}_{\vec{k} \sigma} ={\sigma \over 2}
  \left(J_{0} {\chi_{\rm Mn}\over g\mu_B}- g\mu_B\right).
\end{equation}
Differentiating Eq. (\ref{27}) with respect to $H$, we find:
\begin{equation}
\label{4.10}
\left.{d \mu^B \over d B}\right|_{B=0} = {{\sum_{\vec{k} \sigma}
E^{(1)}_{\vec{k} \sigma} g(E_{\vec{k} \sigma})} \over \sum_{\vec{k}
\sigma} g(E_{\vec{k} \sigma})}
\end{equation}
Substituting Eqs. (\ref{3.10}) and (\ref{4.10}) in Eq. (\ref{5.10}) we
get:
\begin{equation}
\label{6.10}
\chi_{h} = g \mu_B\gamma \left({g\mu_B}-J_{0}{\chi_{\rm Mn}\over
g\mu_B}\right),
\end{equation}
where
\begin{equation}
\label{gamma}
{\gamma}={1 \over {4 N_d}} {[\sum_{\vec{k}\sigma} \sigma g(E_{\vec{k}
	\sigma})]^2 -[\sum_{\vec{k}\sigma} g(E_{\vec{k} \sigma})]^2 \over
	{\sum_{\vec{k}\sigma}g(E_{\vec{k} \sigma})}}.\\
\end{equation}

From Eqs.  (\ref{6.10}) and (\ref{2.10}), we obtain the Mn
susceptibility in the ordered case to be:
\begin{equation}
\label{6.13}
\chi_{\rm Mn}= \beta (g \mu_B)^2{{ (1- J_0 \gamma ) {{B'_S}}(-\beta
J_{0}s_{h}) } \over {1 - \beta J_{0}^2 \gamma {{B'_S}}(-\beta
J_{0}s_{h})}}
\end{equation}
while the hole susceptibility $\chi_{h}$ is:
\begin{equation}
\label{6.13bb}
\chi_{h}= (g \mu_B)^2 \gamma {{1-J_{0}\beta {{B'_S}}(-\beta
J_{0}s_{h})} \over {1 - \beta J_{0}^2 \gamma {{B'_S}}(-\beta
J_{0}s_{h})}}.
\end{equation}

\begin{figure}[t]
\centerline{\includegraphics[width=0.8\columnwidth]{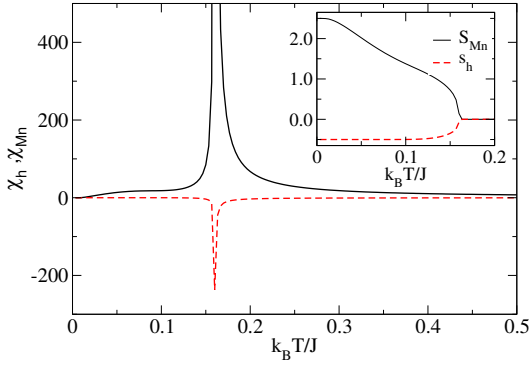}}
\caption{\label{fig1} $\chi_{\rm Mn}(T)$ (full) and $\chi_h(T)$
  (dotted line), for an ordered Mn configuration with $N_{d}=512$,
  $x=0.00926$ and $p=10\%$. The inset shows the corresponding
  magnetizations.}
\end{figure}

These static longitudinal susceptibilities are plotted as a function
of temperature in Fig. \ref{fig1}. As expected, the critical
temperature $T_c$ is marked by a singularity. Since $s_h=0$ for $T\ge
T_c$ and $B'_S(0)=S(S+1)/3$, the denominator in the susceptibilities
gives:
\begin{equation}
\EqLabel{tc} k_BT_c = \gamma J_0^2 {S(S+1)\over 3}
\end{equation}
For $T\ge T_c$, $\sum_{\vec{k}\sigma} \sigma g(E_{\vec{k} \sigma})=0$
(spin degeneracy is restored).  Then [Eq. (\ref{gamma})], $\gamma =
-\sum_{\vec{k}\sigma} g(E_{\vec{k} \sigma})/4N_d =$ $\int_{}^{}dE
\sum_{\vec{k}\sigma}^{}\delta(E-E_{\vec{k}\sigma}) [ -{d \over
dE}f(E)]/4N_d =\rho(E_F)/4n_{\rm Mn}$ (if $k_BT\ll E_F$), where
$\rho(E_F)$ is the density of states per unit volume at the Fermi
energy $E_F$. This value can also be obtained directly from
Eq. (\ref{6.10}), since in the absence of interactions ($J=0$), the
hole susceptibility per unit volume $n_{\rm Mn} \gamma (g\mu_B)^2$
must equal the Pauli susceptibility.

In an effective mass approximation,
$\epsilon_{\vec{k}}=\hbar^2k^2/2m^*$, $\rho(E_F)\sim m^* k_F \sim m^*
(n_h)^{1\over 3}$ $\rightarrow\gamma \sim m^* (px)^{1\over 3}/x$ (the
factors contain only constants). Such an approximation can be used in
two cases: (i) for itinerant models,\cite{Dietl,Konig,Amir} in which
case $m^*=m_h$ is the mass of the heavy hole band, and $J_0
\rightarrow n_{\rm Mn} J_{pd}$ in order to obtain the same
one-electron dispersion (see, {\em e.g.}, Ref. \onlinecite{Konig}). In
this case, we regain the expected $T_c \sim x (px)^{1/3}$ mean-field
scaling with the Mn and hole concentrations $x$ and
$(px)$.\cite{Dietl,Konig,Amir} (ii) for an impurity model on an
ordered lattice, for only nearest neighbor hopping $t$ and $E_F \ll
t$, we have $m^*\sim 1/(ta_L^2)$, where $a_L=a/x^{1\over 3}$ is the
superlattice constant. It then follows that $T_c \sim p^{1/3} J_0^2/t
$. Both $J_0$ and $t$ depend on $x$ through the distance between
neighbors Mn. In Ref. \onlinecite{HF} we showed numerically that at
constant $p$, in the ordered impurity band case, $T_c \sim x$, so one
can infer that here $T_c\sim x p^{1/3}$. In any event, disorder and
thermal fluctuations considerably change these mean-field estimates.

Before discussing disordered systems, it is worth emphasizing why
$\chi_h < 0$. Each hole interacts antiferomagnetically with many Mn
spins, each of which has its magnetization increased by the magnetic
field. This favors an increased polarization of the holes, in a
direction opposite to the applied field. Thus, the direct effect of
the external field on the holes is more than offset by its indirect
effect mediated through exchange with the Mn spins.  As a result, in
the paramagnetic phase $\chi_h$ is strongly enhanced from its
non-interacting, $T$-independent Pauli value. From Eqs. (\ref{6.13})
and (\ref{6.13bb}), we see that
$$ {\chi_h \over \chi_{Pauli}} = {\chi_{Mn} \over \beta (g\mu_B)^2} {
3 - J_0 \beta S(S+1)\over (1- J_0 \gamma)S(S+1)}
$$ This increase can be formally assigned to an enhanced effective
$g$-factor. Consistent with this phenomenology, huge Zeeman shifts
have been measured in both II-VI and III-V DMSs.\cite{Furdyna,Zem}
Spintronic applications based on this high-$T$ effect have been
proposed recently.\cite{MandB,MandB2}

The longitudinal susceptibilities in the disordered case are
calculated similarly. However, we now compute each contribution
$\chi_{\rm Mn}(i) = d S_{\rm Mn}^B(i)/dB|_{B=0}$ and $\chi_h(i) = d
s_{h}^B(i)/dB|_{B=0}$ (for simplicity, we set $g\mu_B=1$, i.e. measure
the susceptibilities in units of $(g\mu_B)^2$). This calculation is
detailed in the Appendix \ref{app1}. We end up with a system of linear
equations for $\chi_{\rm Mn}(i)$ [Eq. (\ref{11.8})]:
$$ \sum_{j} \left[ \delta_{ij} + R_{ij} \right] {\chi_{\rm Mn}}(j)
=\beta(1 -P_{i})B'_S (\beta H_{i})
$$ The matrices $R\sim J^2$ and $P\sim J $ depend only on $B=0$
mean-field quantities [see discussion following Eq. (\ref{11.8})].

It is instructive to compare this result with the ``conventional''
statistical formula for static susceptibility:
\begin{equation}
\EqLabel{suc} \tilde{\chi}_{\rm Mn}={\beta \over N_{d}}
\sum_{ij}\left[\langle \vec{S}_i \vec{S}_j \rangle- \langle \vec{S}_i
\rangle \langle \vec{S}_j \rangle\right]
\end{equation}
At the mean-field level $\langle \vec{S}_i \vec{S}_j \rangle= \langle
\vec{S}_i \rangle \langle \vec{S}_j \rangle$ if $i\ne j$, since the
mean-field density matrix ${\cal D}_0=\exp[-\beta({\cal K}-\mu{\cal
N})]$ is diagonal for different spins [see Eq. (\ref{12})]. It follows
that $ \tilde{\chi}_{\rm Mn}=\sum_{i}\tilde{\chi}_{\rm Mn}(i)/N_d$,
where
$$ {\tilde{\chi}_{\rm Mn}}(i)= \beta \left[\langle \vec{S}_{i}^2
\rangle- \langle \vec{S}_i \rangle^2\right]=\beta B'_S(\beta H_{i}).
$$ This is the solution one obtains if one sets the matrices $R$ and
$P$ to zero, in the full system of linear equations shown
above. Equivalently, comparison with Eq. (\ref{3.88}) shows that this
``conventional'' formula does not account for the contribution from
the supplementary polarization of the holes. Since this is
considerable (singular) near $T_c$, the ``conventional'' expression
gives very wrong results for $T\sim T_c$, although it works well for
$T\rightarrow 0$ or $T\gg T_c$, where the hole susceptibilities are
very small. The reason for this failure is the fact that
Eq. (\ref{suc}) holds when $\langle ... \rangle$ denotes the thermal
average with the exact density matrix, not with the approximate
mean-field density matrix. To be more precise, for a Hamiltonian such
as of Eq. (\ref{1.1}), the total susceptibility of the system is
actually
$$ {\tilde \chi} = {\beta \over N_{d}} \sum_{ij}\left[\langle
(\vec{S}_i+\vec{s}_i)\cdot (\vec{S}_j+\vec{s}_j) \rangle- \langle
\vec{S}_i+\vec{s}_i \rangle \langle \vec{S}_j+\vec{s}_j \rangle\right]
$$ where $\langle ...\rangle$ is the exact thermal average, which can
be evaluated with Monte Carlo simulations. This susceptibility can be
decomposed into a $\tilde{\chi}_{\rm Mn}$ and similar
$\tilde{\chi}_{\rm h}$ as in Eq. (\ref{suc}), but there is also a
cross term containing terms like $\langle \vec{S}_i \vec{s}_j \rangle-
\langle \vec{S}_i \rangle \langle \vec{s}_j\rangle$, which are not
necessarily small near $T_c$. Ignoring these terms, {\em i.e.}
approximating ${\tilde \chi} \approx \tilde{\chi}_{\rm Mn}$ in
Monte-Carlo simulations\cite{MPK2,Sch,Dag} is questionable, especially
when ${\tilde \chi}$ is employed precisely to identify $T_c$.

\begin{figure}[t]
\centering \includegraphics[width=0.8\columnwidth]{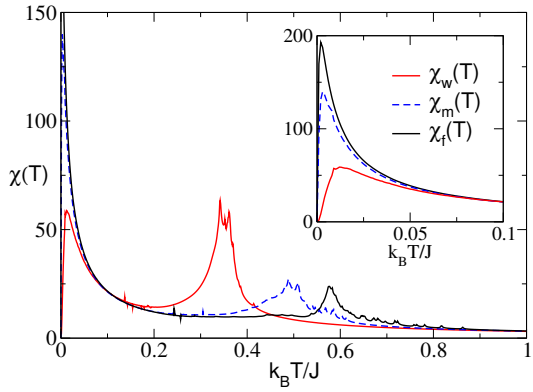}
\caption{{\label{fig2}}Weak, medium and full disorder systems'
  susceptibilities vs. $T$, for $N_{d}=216$, $x=0.00926$ and
  $p=10\%$. Inset focuses on the low-temperature region.}
\end{figure}

We analyze now the effect of disorder on the longitudinal
susceptibility.  We solve numerically the full system of coupled
equations and find $\chi_{\rm Mn}\sim \sum_{i}^{}\chi_{\rm
Mn}(i)$. ($\chi_h$ can be found similarly. However, since $\chi_h\ll
\chi_{\rm Mn}$, it follows that $\chi \approx \chi_{\rm Mn}$). The
results shown in Fig. \ref{fig2} are averaged over 30 disorder
realizations.  We denote by $\chi_{w}(T)$, $\chi_{m}(T)$ and
$\chi_{f}(T)$ the susceptibilities of systems with weak, moderate and
full disorder, as defined in Ref. \onlinecite{MB1}. Unlike the ordered
susceptibility which has only one peak at $T_c$ (see Fig. \ref{fig1}),
the susceptibility of disordered systems has two distinct peaks. One
is at $T_{c}$, while the second peak appears at $T \ll T_c$ and is
seen best in the inset.  With increased disorder, this low-$T$ peak
has increased weight and amplitude and shifts to lower
temperatures. It is due to the weakly-coupled Mn spins which are
positioned far from the regions of the sample where the holes are
located with high probability.\cite{MPK1,HF} These behave like free
spins $\chi_{\rm Mn}(i)\sim 1/T$ except at extremely low temperatures
$k_BT \sim H_i$, where they finally polarize [see Eqs. (\ref{17a}),
(\ref{mf1})].  The high-$T$ peak in $\chi(T)$ marks the mean-field
$T_c$. The high-$T$ peak is determined by the behavior of the
strongly-coupled spins, from the regions where the holes are
located.\cite{MPK1,HF} As observed previously,\cite{MB1,HF,MPK2}
increased disorder leads to higher $T_c$. With increased disorder the
high-$T$ peak also broadens considerably. The explanation is provided
below.

\begin{figure}[t]
\centering \includegraphics[width=0.8\columnwidth]{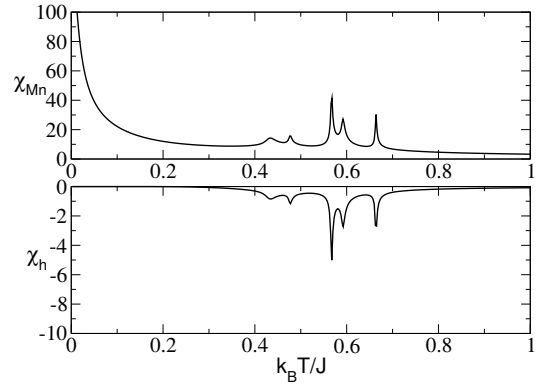}
\caption{{\label{fig3}} $\chi_{\rm Mn}(T)$ and $\chi_h(T)$ for {\em
  single} fully disordered impurity configuration. $N_d=512$,
  $x=0.00924$, $p=0.10$.}
\end{figure}

\begin{figure}[t]
\centering \includegraphics[width=0.8\columnwidth]{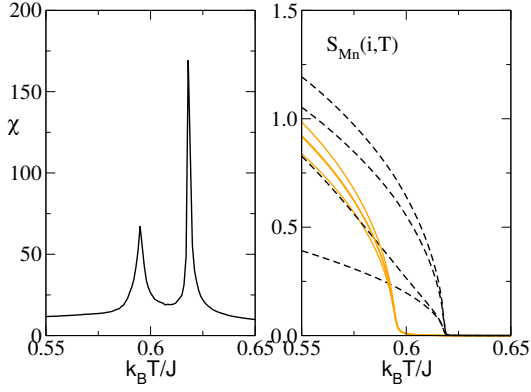}
\caption{{\label{fig4}}Left: total susceptibility of a disordered
  sample, near $T_c$. Right: $S_{\rm Mn}(i,T)$ for spins belonging to
  two different clusters.  $N_d=125$, $x=0.00924$, $p=0.10$.}
\end{figure}

In Fig. \ref{fig3} we show $\chi_{\rm Mn}(T)$ and $\chi_h(T)$ for a
{\em single} disorder realization. Unlike the broad peak near $T_c$ of
the average $\chi_{\rm Mn}$ shown in Fig. \ref{fig2}, here we see
several narrow peaks in a range limited from below by the temperature
where all holes become fully polarized. The highest-$T$ peak occurs at
the $T_c$ of the individual sample. These narrow peaks appear
symmetrically in both susceptibilities (however, $|\chi_{h}| \ll
\chi_{\rm Mn}$).  The number of such peaks and their positions are
different for different samples. Less disordered samples have fewer
peaks, in a smaller temperature range; the number of peaks increases
with system size.\cite{Adel} The origin of these peaks can be inferred
from examining the values of $\chi_{\rm Mn}(i)$ at temperatures where
peaks form. We find that each narrow peak is due to contributions from
a distinct cluster of Mn spins, which are spatially close to one
another and in a region where holes are found with large probability
(these are strongly-coupled spins). The magnetizations $S_{\rm Mn}(i)$
of spins from two different clusters are shown in
Fig. \ref{fig4}. Different clusters polarize at different ``$T_c$'',
depending on their local environment. The local hole spin inside a
cluster also becomes finite at the same $T$.\cite{Adel} From
Fig. \ref{fig4} we see that these ``local $T_c$'' of individual
clusters are the temperatures where $\chi$ has peaks, which thus
signal the establishing of local FM correlations. The average over
many disorder realizations results in a broad peak over the
temperature range where these local FM correlations build up in the
system. This range is larger for more disorder, which implies more
inhomogeneity. On the other hand, it decreases with increasing
$x$,\cite{Adel} since for higher $x$ fluctuations in the local
concentration are reduced.

Strictly speaking, the true $T_c$ of a sample is not related to this
mean-field $T_c$ estimate where various strongly-coupled clusters
begin to polarize. Instead, several magnetized clusters must appear
all throughout the sample, and correlations between their
magnetizations must be established (through exchange of polarized
holes) before long-range magnetic order develops. This is
qualitatively like the picture of magnetic polarons (each polarized
cluster represents a magnetic polaron),\cite{dasS} except that there
$T_c$ is marked by percolation of the growing polarons, since the
polarons cannot exchange polarized holes.  This susceptibility peak at
the mean-field $T_c$ therefore denotes the characteristic temperature
where the clusters (magnetic polarons, or local FM correlations) begin
to form, also denoted by $T^*$ in other studies.\cite{Dag,Mayr} Monte
Carlo simulations are needed to determine $T_c$ for the phase
transition to long-range magnetic order.

\section{The Random-Phase Approximation}

We now consider the magnetic response to a general time-dependent
magnetic field :
\begin{equation}
\EqLabel{bb} \vec{B}(i,t) = e^{\eta t} \int_{-\infty}^{\infty}{d\omega
\over 2\pi} {1 \over V} \sum_{\vec{q}}^{} e^{i(\vec{q}\cdot
\vec{R}_i-\omega t)} \vec{B}(\vec{q},\omega)
\end{equation}
turned on adiabatically at $t=-\infty$ ($\eta>0$ is infinitesimally
small).  The proper equations of motion for any operator ${\cal A}(t)$
and for the density matrix ${\cal D}(t)$ are obtained from the
minimization of the action:\cite{Ripka}
\begin{equation}
\nonumber S_{\cal{H}}=-\int_{t_i}^{t_f} {d t} \, Tr \left \{ {\cal
A}\hbar{d {\cal D} \over d t}+i {\cal A} [{\cal H},{\cal D}] \right \}
+ Tr{\cal D}(t_f){\cal A}(t_f),
\end{equation}
subject to the constraints $\delta{\cal D}(t_i)=\delta {\cal
A}(t_{f})=0$. Here, ${\cal H}(t)$ is the full Hamiltonian of the
system [Eq. (\ref{1.1}) in this case]. Approximation schemes are
obtained by solving $\delta S_{\cal{H}}= 0$ for various classes of
trial density matrices.

For a time-dependent mean-field calculation, the trial density matrix
${\cal{D}} (t)={e^{-\beta [{\cal K}(t)- \mu {\cal N}]}}/{{\cal Z}(t)}$
is defined by a variational quadratic Hamiltonian ${\cal K}(t)$ and
satisfies the initial condition $ {\cal{D}} (-\infty)= {\cal D}_{0}$,
where ${\cal D}_{0}$ is the mean-field density matrix of the
unperturbed system.  Like for the static mean-field derivation, we
define the expectation values [compare with Eqs. (\ref{16}),
(\ref{17})]:
\begin{eqnarray}
\label{7.299}
&\rho_{i\alpha, j\beta}(t)= Tr \{ {\cal D}(t) c^{\dagger}_{j\beta}
c_{i\alpha}\} \\
\label{7.3}
&\vec{S}(i,t)=Tr\{{\cal D}(t) \vec{S}_{i}\}.
\end{eqnarray}
${\cal{A}}$ is taken to have a general quadratic dependence:
\begin{equation}
\label{88.3}
{\cal{A}}(t)=\sum_{i, j \atop \alpha\beta} a_{i\alpha, j \beta}(t)^{}
c^{\dagger}_{i\alpha} c_{j\beta}-\sum_{i}\vec{A_i}(t) \cdot \vec{S_i}.
\end{equation}

In terms of these quantities, $S_{\cal{H}}$ becomes:
\begin{widetext}
\begin{eqnarray}
S_{{\cal H}} \!&=& \!-\hbar \!\!\int_{t_i}^{t_f}\!\!\! \!\! d t
\!\Big[\! \sum_{i j, \alpha\beta}\!\!a_{i\alpha, j \beta}(t) {d
\rho_{j\beta,i\alpha}(t)\over dt} \!-\!\!\sum_{i}\vec{A_i}(t)
\!\cdot\! {d \vec{S}(i,t) \over dt} \Big] \! -i \!\!\int_{t_i}^{t_f}
\!\!\! \!\! d t \!\!\! \!\! \sum_{i\alpha, j\beta, k\lambda}\!\!\!
\!\!  h_{i\alpha, j\beta}(t) \Big[ \rho_{j\beta,k\lambda}(t)
a_{k\lambda,i\alpha}(t)-
a_{j\beta,k\lambda}(t)\rho_{k\lambda,i\alpha}(t)\Big] \nonumber \\ &+&
\! \hbar \int_{t_i}^{t_f}\!\!\! \!\! {d t}^{} \sum_{i}\vec S(i,t)
\cdot \Big[\vec{A_i}(t) \times \vec{H}_i(t) \Big] +\sum_{i\alpha,
j\beta} a_{i\alpha,j\beta}(t_f)\rho_{j\beta,i\alpha}(t_f)-\sum_{i}
\vec A_{i}(t_f) \cdot \vec S(i,t_f) \nonumber
\end{eqnarray}
\end{widetext}
where [compare with Eqs. (\ref{mf1}), (\ref{mf2})]:
\begin{equation}
\label{9.3}
h_{i\alpha,j\beta}(t)=t_{ij} \delta_{\alpha \beta}+\delta_{ij}{
  \vec{\sigma}_{\alpha\beta}\over 2}\cdot \Big[\sum_{k} J_{ik} \vec
  S(k,t) - g \mu_B \vec{B}(i,t)\Big]
\end{equation}
\begin{equation}
\label{12.3}
\vec {H}_i(t)= g \mu_B \vec{B}(i,t) -\sum_{j\alpha\beta}
J_{ij}{\vec\sigma_{\alpha\beta} \over 2}\rho_{j\beta,j\alpha}(t)
\end{equation}
From $\delta {\cal S}_{{\cal H}}/\delta a_{i\alpha, j\beta} =0$, where
$\delta a_{i\alpha, j\beta}(t_f) =0$, we obtain the equation of motion
for $\rho_{j\beta,i\alpha}(t)$:
\begin{equation}
\label{10.3}
i \hbar {d\rho_{j\beta,i\alpha}(t) \over dt}
\!=\!\!\sum_{k\lambda}\!\left[ h_{j\beta,k\lambda}(t)^{}
\rho_{k\lambda,i\alpha}(t)\!-\!  \rho_{j\beta,k\lambda}(t)^{}
h_{k\lambda,i\alpha}(t)\right]
\end{equation}
which is just the matrix form of the expected $i\hbar d\rho/dt = [H,
\rho]$. Similarly, the condition $\delta {\cal S}_{{\cal H}}/\delta
\vec{A}_i(t) =0$, where $\delta \vec{A}_i(t_f) =0$, leads to the
expected equation of motion:
\begin{equation}
\label{12.3a}
 {d \over dt} \vec S(i,t) = -\vec {H}_i(t) \times \vec S(i,t).
\end{equation}
Eqs. (\ref{10.3}) and (\ref{12.3a}) describe the general time
evolution for any value of the external field. We are interested in
the linear regime of a perturbationally small external field. As a
result, we need to solve Eqs. (\ref{10.3}) and (\ref{12.3a}) to first
order. We introduce the notation [see Eqs. (\ref{16}), (\ref{17})]:
\begin{eqnarray}
\nonumber \rho_{j\beta,i\alpha}(t)&= &\delta_{\alpha\beta}
\rho_{ji,\alpha}+ \delta\rho_{j\beta,i\alpha}(t) +\cdots \\\nonumber
\vec S(i,t)&= &{\vec e_z} S_{\rm Mn}(i)+ \delta\vec S(i,t)+\cdots
\end{eqnarray}
where we use the convention that all quantities denoted $\delta X$
depend linearly on the external magnetic field $\vec{B}(i,t)$. To
first order, the effective Hamiltonian (\ref{9.3}) and the effective
magnetic field (\ref{12.3}) become [Eqs. (\ref{mf1}), (\ref{mf2})]:
\begin{eqnarray}
\nonumber
h_{i\alpha,j\beta}(t)&=&\delta_{\alpha\beta}^{}h_{ij,\alpha}+\delta_{ij}
\delta h_{i, \alpha\beta}(t)+\cdots\\ \nonumber \vec{H}_i(t) &=& {H}_i
\vec{e}_z + \delta \vec{H}_i(t) +\cdots
\end{eqnarray}
where
\begin{eqnarray}
\nonumber &\delta h_{i, \alpha \beta}(t)= {
 \vec{\sigma}_{\alpha\beta}\over 2}\cdot \Big[\sum_{k} J_{ik}
 \delta\vec S(k,t) - g \mu_B \vec{B}(i,t)\Big] \\\label{dHe}
 &\delta\vec {H}_i(t)= g \mu_B \vec{B}(i,t) -\sum_{j\alpha\beta}
 J_{ij}{\vec\sigma_{\alpha\beta} \over
 2}\delta\rho_{j\beta,j\alpha}(t)
\end{eqnarray}

We now substitute these expressions into Eqs. (\ref{10.3}) and
(\ref{12.3a}). The zero order (static) terms cancel out. After a
time-domain Fourier transformation (taking into consideration the
adiabatic term $e^{\eta t}$), we obtain:
\begin{eqnarray}
\label{4.4}
&\hbar(\omega +i \eta) \delta \rho_{j \beta, i \alpha}(\omega) =
 [\delta h_{ j, \beta \alpha}(\omega) \rho_{ji,\alpha}-\delta h_{ i,
 \beta \alpha}(\omega) \rho_{ji,\beta}]\nonumber \\ & + \sum_{k}
 [h_{jk,\beta}^{} \delta \rho_{k\beta,i\alpha}(\omega)-h_{ki,\alpha}
 \delta \rho_{j\beta,k\alpha}(\omega)] ,
\end{eqnarray}
\begin{equation}
\label{4.4a}
i (\omega +i \eta) \delta \vec{S}(i,\omega)= \vec{e_z} \times \Big[
H_i \delta \vec{S}(i,\omega)- \delta \vec{H}_i(\omega) S_{\rm
Mn}(i)\Big]
\end{equation}
Let us consider first Eq. (\ref{4.4a}) -- after all, we are interested
in the linear change $\sum_{i}^{} \delta \vec{S}(i,t)$ of the
magnetization, related to the dynamic susceptibility.
Eq. (\ref{4.4a}) projects into two different equations for the
components $\delta S_z(i,\omega)$ and $\delta S_+(i,\omega)= \delta
S_{x}(i,\omega) +i \delta S_{y}(i,\omega) $:
\begin{eqnarray}
 \nonumber & (\omega + i \eta) \delta S_{z}(i,\omega)=0,\\
\label{8.4} &( \omega - H_{i} + i \eta)  \delta S_+(i,\omega)= -S_{\rm
  Mn}(i) \delta H^+_i(\omega)
\end{eqnarray}
The first equation shows that for $\omega \ne 0$, $\delta
S_{z}(i,\omega)=0$ and therefore $S_{\rm Mn}(i)$ is conserved (the
static case $\omega =0$ was investigated in the previous section). It
follows that we can only define a transverse dynamical
susceptibility. To calculate it, we need the values of $\delta
H^+_i(\omega)$, which depend on $\delta \rho_{j \downarrow, j \uparrow
}(\omega)$ [see Eq. (\ref{dHe})]. In turn, these depend on all $\delta
\rho_{j \downarrow, k \uparrow }(\omega)$ components [see
Eq. (\ref{4.4})]. Instead of working directly with these, it is more
convenient to introduce:
\begin{equation}
\label{1.5}
X_{nm}(\omega)=\sum_{ij}\psi^*_{m\downarrow}(j) \delta
\rho_{j\downarrow,i\uparrow}(\omega)\psi_{n\uparrow}(i)
\end{equation}
where $\psi_{n\sigma}(i)$ are the self-consistent mean-field
eigenfunctions of Eq. (\ref{25}) (here $B=0$, since there is no static
field applied) which are orthonormal and complete.  Eqs. (\ref{8.4})
and (\ref{4.4}) (for $\beta =\downarrow, \alpha =\uparrow$) now
become:
\begin{widetext}
\begin{equation}
\label{4.5}
(\omega -H_i +i\eta) \delta S_+(i,\omega) = 2S_{\rm Mn}(i)
  \sum_{nm}^{}J_{n\uparrow, m\downarrow}(i) X_{nm}(\omega) - g \mu_B
  S_{\rm Mn}(i) B_+(i,\omega)
\end{equation}
\begin{equation}
\EqLabel{2.55} (\hbar\omega +E_{n\uparrow} - E_{m\downarrow} + i \hbar
\eta) X_{nm}(\omega) = \left[f(E_{n\uparrow}) - f(E_{m\downarrow})
\right] \left[ \sum_{i}^{} J_{m\downarrow, n\uparrow}(i) \delta
S_+(i\omega) - g \mu_B \sum_{i}^{} \psi^*_{m\downarrow}(i)
\psi_{n\uparrow}(i) B_+(i,\omega)\right]
\end{equation}
\end{widetext}
where $J_{n\alpha, m\beta}(i)={1 \over 2} \sum_{j} J_{ij} \psi^{*}_{n
\alpha}(j) \psi_{m \beta}(j)$, the eigenenergies $E_{n\sigma}$ and the
occupation numbers $f(E_{n\sigma})$ depend on known static mean-field
quantities.  (Note: for $\omega \ne 0$, the chemical potential $\mu$
remains unchanged to its static self-consistent mean-field value, to
first order, since from Eq. (\ref{4.4}) it follows immediately that $
(\hbar \omega + i\eta)\sum_{i,\alpha}^{} \delta \rho_{i \alpha, i
\alpha} =0 $).

The $N_d + N_d^2$ Eqs. (\ref{4.5}) and (\ref{2.55}) [or alternatively,
  Eqs. (\ref{4.4}) and (\ref{4.4a})] are the generalized Random Phase
  Approximations (RPA) equations at finite temperature. In the limit
  $T=0$, they indeed reduce to the RPA equations derived in
  Ref.~\onlinecite{RPA}. They can be replaced by a system of only
  $N_d$ linear equations for $\delta S_+(i,\omega) $ by substituting
  the $X_{nm}(\omega)$ variables from (\ref{2.55}) into
  (\ref{4.5}). The final result is:
\begin{equation}
\label{5.5a}
\sum_{j}M_{ij}(\omega)\delta S_+(j,\omega)=b_{i}(\omega)
\end{equation}
where
\begin{widetext}
\begin{equation}
\label{6.5}
  M_{ij}(\omega)=\delta_{ij}( \omega-H_{i}+i\eta) -2S_{\rm Mn}(i)
\sum_{n,m} {J_{n\uparrow,m\downarrow}(i)J_{m\downarrow,n\uparrow}(j)
[f(E_{n\uparrow})-f(E_{m\downarrow})] \over \hbar \omega +
E_{n\uparrow} - E_{m\downarrow} +i\hbar\eta}
\end{equation}
\begin{equation}
\label{7.5}
b_i(\omega) = -g \mu_B S_{\rm Mn}(i) \left[B_+(i,\omega) +
 2\sum_{n,m}^{}
 J_{n\uparrow,m\downarrow}(i){f(E_{n\uparrow})-f(E_{m\downarrow})\over
 \hbar \omega + E_{n\uparrow} - E_{m\downarrow} +i\hbar\eta}
 \sum_{j}^{} \psi^*_{m\downarrow}(j) \psi_{n\uparrow}(j)
 B_+(j,\omega)\right]
\end{equation}
\end{widetext}

\section{The transverse dynamical susceptibility}

The transverse dynamical susceptibility is defined as:
\begin{equation}
\EqLabel{ss} \delta S_+(\vec{q}, \omega) = { n_{\rm Mn} \over N_d}
\sum_{i}^{}e^{-i\vec{q}\cdot \vec{R}_i} \delta S_+(i, \omega) =
{\chi(\vec{q}, \omega)\over g\mu_B} B_+(\vec{q}, \omega)
\end{equation}
for each transverse component of the applied field $B_+(i, \omega) =
e^{i\vec{q}\cdot \vec{R}_i} {B}_+(\vec{q}, \omega)/V$ [see
Eq. (\ref{bb})]. Note that all $b_i(\omega) \sim B_+(\vec{q}, \omega)$
[Eq. (\ref{7.5})] and thus all $\delta S_+(i, \omega)$ of
Eq. (\ref{5.5a}) are indeed proportional to $B_+(\vec{q}, \omega)$.
In other words, for each $\vec{q}$ and $\omega$ of interest, we can
set $B_+(\vec{q}, \omega)=1 \rightarrow B_+(i, \omega) =
e^{i\vec{q}\cdot \vec{R}_i}/V$. We can then compute the matrix
elements $M_{ij}(\omega)$ and $b_{i}(\omega)$ for any finite
temperature, and solve Eq. (\ref{5.5a}) for $\delta
S_+(i,\omega)$. With this convention, the transverse susceptibility
per unit volume is $\chi(\vec{q}, \omega) = g \mu_B n_{\rm Mn}/N_d
\sum_{i}^{} \exp{(-i\vec{q}\cdot \vec{R}_i )}\delta S_+(i,\omega)$.

In the ordered case, this calculation can be carried out explicitly
using Eqs. (\ref{4.66}), (\ref{5.66}) and (\ref{6.66}). The result is:
\begin{equation}
\EqLabel{15.6} \chi(\vec{q},\omega) = { (g\mu_B)^2 n_{\rm Mn}S_{\rm
Mn} \left[1 - J_{\vec{q}}F_{\vec{q}}(\omega) \right] \over \hbar
\omega +J_{0} s_{h}+ {1 \over 2} S_{\rm Mn}|J_{\vec{q}}|^2
F_{\vec{q}}(\omega) }
\end{equation}
where $J_{\vec{q}} = \sum_{\vec{\delta}}^{} e^{i \vec{q}\cdot
  \vec{\delta}}J_{\vec{\delta}}$ [$ J_{\vec{\delta}}= J_{ij}$ for
  which $\vec{\delta}= \vec{R}_i - \vec{R}_j$].  $F_{\vec{q}}(\omega)
  \!=\!  {1 \over N_d} \!\sum_{k}^{} [
  f(\!E_{\vec{k},\downarrow}\!)\!-\!f(\!E_{\vec{k}-\vec{q},\uparrow}\!)]/
  [ \hbar \omega \!+\!  \epsilon_{\vec{k}-\vec{q}} -
  \!\epsilon_{\vec{k}} \!+ \!J_0 S_{\rm Mn}] $ is the spin-polarized
  electron-hole ``bubble'' expected to appear in RPA-level
  approximations.

\begin{figure}[b]
\centering \includegraphics[width=0.8\columnwidth]{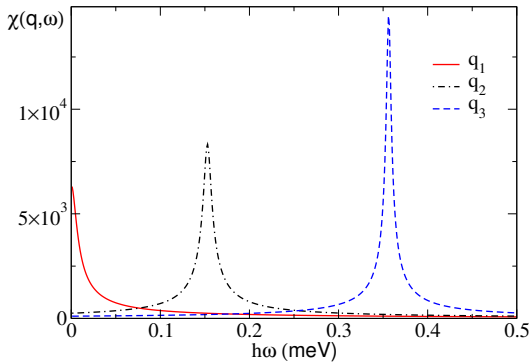}
\caption{{\label{fig5}} $\chi({\vec{q},\omega})$ for the ordered
case. $\vec{q}_1=0$, $\vec{q}_2={\pi\over 15a} (2,0,0)$ and
$\vec{q}_3= 2\vec{q}_2$.  $N_{d}=125$, $x=0.00924$, $p=0.1$, $T=0$.}
\end{figure}

\begin{figure}[t]
\centering \includegraphics[width=0.8\columnwidth]{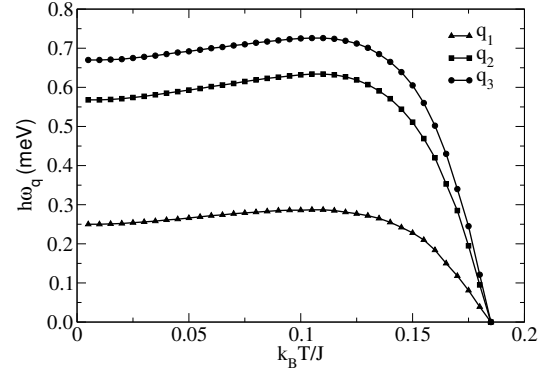}
\caption{{\label{fig6}} Spin-wave energies $\hbar \omega_{\vec{q}}(T)$
  for $\vec{q}_1={\pi\over 9a}(1,1,0)$, $\vec{q}_2=2\vec{q}_1$ and
  $\vec{q}_3=3\vec{q}_1$; $N_{d}=216$, $x=0.00924$, $p=0.1$. }.
\end{figure}

Singularities in $\chi(\vec{q},\omega)$ mark the spectrum
$\hbar\omega_{\vec{q}}$ of the spin-wave modes. In Fig. \ref{fig5}, we
plot $\chi(\vec{q}, \omega)$ for different $\vec{q}$, at $T=0$. The
values $\hbar\omega_{\vec{q}}$ of the singularities indeed coincide
with the spin-wave spectra of Ref. \onlinecite{RPA}. At finite-$T$,
$\chi(\vec{q},\omega)$ for the ordered system still has a single peak,
but its energy first increases, then decreases with $T$. This behavior
is generic, as shown in Fig. \ref{fig6}. Such non-monotonic behavior
is easy to understand, since [see denominator of Eq. (\ref{15.6})]
$\hbar\omega_{\vec{q}} = J_0 |s_h| - {1 \over 2} S_{\rm
Mn}|J_{\vec{q}}|^2 F_{\vec{q}}(\omega_{\vec{q}})$. At low-$T$, $|s_h|$
is constant while $S_{\rm Mn}$ decreases with $T$ (see inset of
Fig. \ref{fig1}), and $\hbar\omega_{\vec{q}}$ increases. Once the
holes start to depolarize, $\hbar\omega_{\vec{q}}\rightarrow 0$. The
exception is the case $\vec{q}=0$, where one finds
$\chi(\vec{0},\omega)=(g\mu_B)^2 n_{\rm Mn}S_{\rm
Mn}/(\hbar\omega)$. The $\omega=0$ singularity signals the Goldstone
boson at all $T< T_c$ where one can define a transversal
susceptibility.

\begin{figure}[t]
\centering \includegraphics[width=0.8\columnwidth]{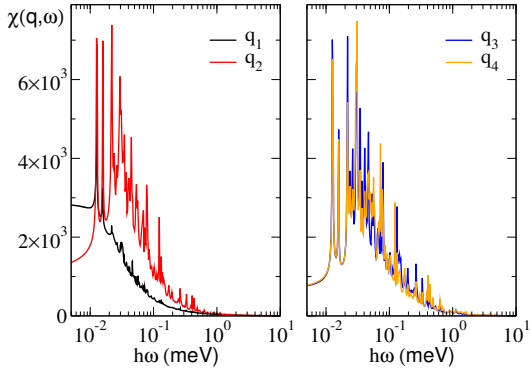}
\caption{{\label{fig7}} $\chi(\vec{q},\omega)$ for an {\em individual}
  fully-disordered configuration, for $\vec{q}_1 =0$, $\vec{q}_2={\pi
  \over 15a}(4,0,0)$, $\vec{q}_3={\pi \over 15a}(8,0,0)$ and
  $\vec{q}_4 = {\pi \over 15a}(8,8,8)$. $N_d=125$, $x=0.00924$ and
  $p=0.1$, $T=0$.}
\end{figure}

In Fig. \ref{fig7}, we show $\chi(\vec{q},\omega)$ of a {\em single}
fully-disordered configuration, for four different values of
$\vec{q}$. In contrast to the ordered case (Fig. \ref{fig5}), here we
see multiple peaks in $\chi(\vec{q},\omega)$, which appear at the same
energies for all values of $\vec{q}$. The explanation is that disorder
breaks translational invariance and $\vec{q}$ is no longer a good
quantum number. As discussed in Ref. \onlinecite{RPA}, in this case
many spin-wave modes become localized and even the extended modes do
not carry well defined momentum. As a result, an external magnetic
field $B_+(\vec{q}, \omega)$ can excite all the spin-waves of energy
close to $\omega$ (conservation of energy) since conservation of
momentum no longer holds. The narrow peaks in $\chi(\vec{q},\omega)$
for various $\vec{q}$ occur at the same frequencies because they
couple to the same spin-wave modes. The differences are mostly in the
amplitude of the peaks, but even these converge and
$\chi(\vec{q},\omega)$ becomes roughly independent of $\vec{q}$ for
moderate to large $\vec{q}$-values.

\begin{figure}[b]
\centering \includegraphics[width=0.8\columnwidth]{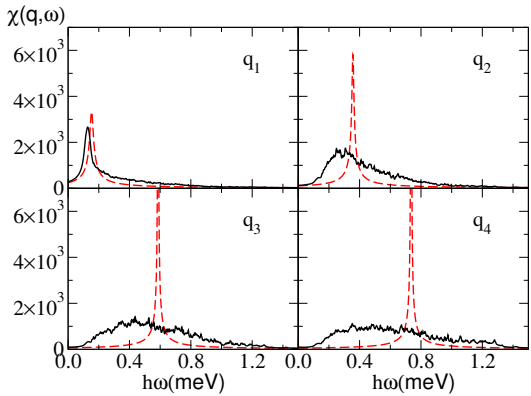}
\caption{{\label{fig8}} $\chi(\vec{q},\omega)$ for ordered (dashed)
and weakly disordered (full) systems. $T=0$, $x=0.00924$, $p=0.10$,
$N_d=125$, and $\vec{q}_{1}={\pi\over 15a} (2,0,0)$,
$\vec{q}_2=2\vec{q}_1$, $\vec{q}_{3}={\pi\over 15a} (4,4,0)$,
$\vec{q}_{4}={\pi\over 15a} (4,4,4)$. }
\end{figure}

To restore invariance to translations, we have to average over all
disorder realizations. The arguments presented above suggest that the
averaged susceptibility is very different from that of an ordered
system even for weak disorder.  This is indeed confirmed in
Fig. \ref{fig8}, where we compare $\chi(\vec{q},\omega)$ for ordered
and weakly-disordered systems (averaged over 15 disorder
configurations). Short wavevectors ($\vec{q}_1$) probe long
length-scales, i.e.  the extended modes. For weak disorder, these are
not strongly perturbed and the results are fairly similar. However, at
short wavelengths ($\vec{q}_2$, $\vec{q}_3$ and $\vec{q}_4$), the
response in the disordered system is dominated by the localized modes
and leads to a roughly $\vec{q}$-independent, extremely broad peak in
the dynamic susceptibility. Note that even the extended modes, which
occupy the center of the spectrum,\cite{RPA} contain some short
wavelength contributions and thus are probed by fields with large
$\vec{q}$.

\begin{figure}[t]
\centering \includegraphics[width=0.8\columnwidth]{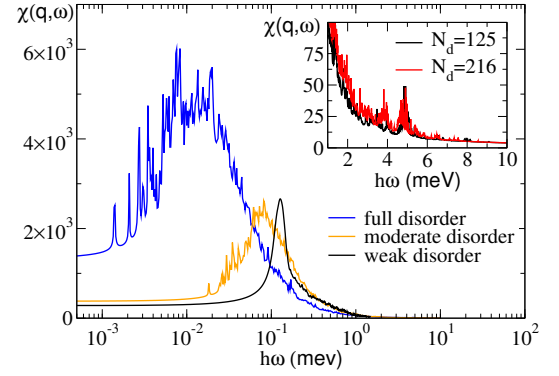}
\caption{{\label{fig9}} $\chi(\vec{q},\omega)$ for ordered (dashed)
and weakly disordered (full) systems, for $\vec{q}={\pi\over 15a}
(2,0,0)$, $T=0$, $x=0.00924$, $p=0.10$, $N_d=125$. Inset: the
high-energy tail for fully disordered configurations, with
$\vec{q}={\pi\over 15a} (4,0,0)$ ($N_d=125$) and $\vec{q}={\pi\over
18a} (4,0,0)$ ($N_d=125$).}
\end{figure}

The change is even more drastic if disorder is increased and more
modes become localized. A comparison for $\chi(\vec{q},\omega)$ for
different levels of disorder, averaged over 20 disorder realizations,
are shown in Fig. \ref{fig9} on a logarithmic scale. The curves are
not yet smooth, meaning that one needs to average over more
samples. However, this is time consuming and the main features are
already apparent. With increased disorder, the peaks become broader
and shift towards lower energies. This is consistent with
Ref. \onlinecite{RPA}, which found an increased density of localized
spin-waves at lower energies in the more disordered systems. Curves in
Fig. \ref{fig9} saturate to a finite value as $\omega \rightarrow 0$
because we used a finite value for $\eta$. From analyzing the
dependence on $\eta$, we find\cite{Adel} that
$\chi(\vec{q},\omega)\sim 1/\omega$ as $\omega\rightarrow 0$ for all
$\vec{q}$.  This is the expected result: if one applies a static
magnetic field transverse to the direction of the magnetization, the
magnetization axis will rotate to become parallel to the applied
field. This is a finite change in magnetization no matter how small
the applied field is, and the static transverse susceptibility is
infinite. In the ordered system, the momentum conservation prevents
this singularity from being observed unless $\vec{q}=0$.

The high-energy tail of $\chi(\vec{q},\omega)$ is shown in the inset
of Fig.  \ref{fig9}, for two different system sizes. The $N_d=125/216$
curve has been been averaged over 40/20 disorder realizations. Because
of the different system sizes, we have to investigate different
vectors in the Brillouin zone. However, as already emphasized, here
$\chi(\vec{q},\omega)$ is roughly independent of $\vec{q}$. Indeed,
the two curves are very similar, suggesting also that finite-size
effects are negligible. As discussed in Ref. \onlinecite{RPA}, the
high-energy collective modes are spin-waves localized inside
strongly-interacting clusters (magnetic polarons). In particular, the
peak at $\sim 5 $meV is due to clusters made of two nearest-neighbor
Mn.\cite{RPA}

On a technical note, we used $\eta=0.02/0.05 J$ for ordered/disordered
systems. A finite $\eta$ implies a finite spin-wave lifetime, due to
scattering on other spin-waves (neglected at the RPA level) and is
necessary to avoid singularities in numerical computations. Also, all
susceptibilities shown are in units of $(g\mu_B)^2 n_{\rm Mn}$.

\section{Conclusions}

In previous work, we showed that positional disorder strongly
influences the shape of the magnetization curve. Due to disorder, some
of the charge carriers are localized in regions of the sample which
have a large local Mn concentration. These regions (clusters or
polarons) polarize at much higher temperatures than an ordered sample
does, due to the much stronger effective coupling between the spins
and holes. Charge carriers delocalized amongst several clusters help
correlate their magnetizations at somewhat lower temperatures, thus
leading to the appearance of long-range magnetic order. The concave
magnetization curves we obtain\cite{HF} are in qualitative agreement
with those obtained by other studies\cite{dasS3} which explicitly take
into account the positional disorder, especially for low charge
carrier densities.  The large inhomogeneity induced by the localized
states was also shown to influence the spin-wave spectrum, leading to
localized modes at both low and high energies.\cite{RPA} The
low-energy localized spin-waves are, in fact, spin-flips of the weakly
interacting spins, whereas the high-energy localized modes are
spin-flips inside strongly coupled clusters.

Here, we show that the magnetic susceptibilities are also strongly
sensitive to positional disorder. In particular, even very little
disorder leads to a qualitatively different behavior of
$\chi(\vec{q},\omega)$ compared to the ordered case, as shown in
Fig. (\ref{fig8}): instead of a Lorentzian centered at a well-defined
spin-wave frequency $\hbar\omega_{\vec{q}}$, in the disordered case we
obtain a very broad, roughly $\vec{q}$-independent peak, which extends
over the entire range of the spin-wave spectrum. In the traditional
weakly-scattering case, the average over all disorder realizations
leads to a finite lifetime of the excitations, but momentum is still a good
quantum number. By contrast, here even small amounts of disorder
induce localization of some of the charge-carriers,\cite{HF} which in
turns leads to localization of some of the spin-wave modes.\cite{RPA}
Dealing with localization is well beyond the realm of applicability of
weak-scattering arguments. Indeed, as we show here, the susceptibility
in the presence of disorder is not just like that of an ordered
system, but with a finite life-time; instead, at any given $\vec{q}$ a
transversal field can couple to all the spin-waves in the system, and
therefore $\chi(\vec{q}, \omega)$ is finite for all $\omega$ in the
spin-wave spectrum. The only ingredient necessary for this dramatic
change in the shape of $\chi(\vec{q}, \omega)$ is the existence of
some charge carrier localized states. On general grounds, one expects
that to be the case at all $x$ below and near the MIT. This prediction
could be confirmed once neutron scattering experiments are performed
on DMS.

The formalism we developed here can be trivially extended to more
complicated cases, for instance to include anisotropies due to strain
or spin-orbit coupling, non-collinear self-consistent ground-states or
other supplementary terms such as on-site disorder, electron-electron
interactions, etc. It is very unlikely that any such extra terms can
completely inhibit the appearance of localization. As a result, their
addition can only lead to some quantitative changes, but qualitatively
the susceptibilities behave as the ones we derived using this simple
impurity-band model.

{\bf Acknowledgements:} This work was supported by NSERC of Canada and
by the Research Corporation.

\appendix*

\section{Static Longitudinal Susceptibility}
\label{app1}

In this appendix we sketch the derivation of the static longitudinal
susceptibility in the disordered case. Combining Eqs. (\ref{17a}) and
(\ref{mf1}), we find $ S^{B}_{\rm Mn}(i) = B_{S}[\beta( B-\sum_{j}
J_{ij} s^{B}_{h}(j))], $ (we set $g\mu_B=1$ for simplicity). Then,
\begin{equation}
\label{3.88}
{\chi_{\rm Mn}}(i)= \left.{d S^B_{\rm Mn}(i)\over dB}\right|_{B=0}=
  \beta [1-\sum_{j} J_{ij}\chi_{h}(j)] {{B'_S}}(\beta H_{i})
\end{equation}
where ${B'_S}(x)= {d / dx}{{B_{S}}}(x)$.  Let us now compute
$\chi_{h}(i)$. From Eq. (\ref{mf2}) we find, to first order in $B$,
that
\begin{equation}
h^{B}_{ij,\sigma}=h_{ij,\sigma}+ \delta_{ij} {\sigma \over 2}
[\sum_{k} J_{ik} \chi_{\rm Mn}(k)-1 ] B+\cdots
\end{equation}
and therefore [Eq. (\ref{24})] ${\cal K}_{el}^B = {\cal K}_{el} + B
  {\cal V}_{ex} +\cdots$, where:
$$ {\cal V}_{ex} = \sum_{i,\sigma}^{} {\sigma \over 2} [\sum_{j}
J_{ij} \chi_{\rm Mn}(j)-1 ] c^{\dag}_{i\sigma} c_{i\sigma}
$$ We now use perturbation theory to find $E^{B}_{n\sigma}$ and
$\psi^{B}_{n \sigma}(i)$ to first order in $B$. In a disordered system
all degeneracies are lifted, and thus $E^{B}_{n \sigma}=E_{n
\sigma}+E^{(1)}_{n \sigma}B + \cdots$, $\psi^{B}_{n \sigma}(i)=\psi_{n
\sigma}(i)+\psi^{(1)}_{n \sigma}(i) B +\cdots$, where
\begin{eqnarray}
\label{en1} &E^{(1)}_{n \sigma}\!=\!\langle \psi_{n \sigma}|{\cal V}_{ex}| \psi_{n
\sigma} \!\rangle \!=\! \sigma\!\!\sum_{j} \!J_{n \sigma, n \sigma}(j)
\chi_{\rm Mn}(j)\!-\!{\sigma \over 2 }\\ &\psi^{(1)}_{n \sigma}(i)=
\sum_{m \ne n} {{\langle \psi_{m \sigma}|{\cal V}_{ex}| \psi_{n
\sigma} \rangle} \over E_{n\sigma}-E_{m\sigma}} \psi_{m \sigma}(i)
\nonumber\\
\label{fi1} &= {\sigma} \sum_{m \ne
n, j} {{ J_{m \sigma, n \sigma}(j) \chi_{\rm Mn}(j)} \over
E_{n\sigma}-E_{m\sigma}}\psi_{m \sigma}(i)
\end{eqnarray}
where $J_{n\alpha, m\beta}(i)={1 \over 2} \sum_{j} J_{ij} \psi^{*}_{n
\alpha}(j) \psi_{m \beta}(j)$.  Finally, differentiating
Eq. (\ref{27}), we find:
\begin{equation}
\label{8.88}
\left.{d \mu^B \over d B}\right|_{B=0} = {{\sum_{n \sigma} E^{(1)}_{n
 \sigma} g(E_{n \sigma})} \over \sum_{n \sigma} g(E_{n \sigma})}.
\end{equation}

Since $s^{B}_{h}(i)= {1 \over 2}\sum_{n\sigma}\sigma
|\psi^{B}_{n\sigma}(i)|^{2} f(E^{B}_{n\sigma})$, it follows that
\begin{eqnarray}
\nonumber & \chi_{h}(i)=\sum_{n\sigma} {\sigma\over 2}\left\{
\left[\psi^{(1)}_{n\sigma}(i) \psi^{*}_{n\sigma}(i) + c.c. \right]
f(E_{n\sigma}) \right.\\
\label{9.88} & + \left.|\psi_{n\sigma}(i)|^2 \beta \left(E^{(1)}_{n \sigma}
-{d \mu \over d H}\right) g(E_{n\sigma}) \right\}.
\end{eqnarray}
Substituting the expressions for $E^{(1)}_{n \sigma}$, $\psi^{(1)}_{n
\sigma}(i)$ and ${d \mu^B \over d B}|_{B=0}$ from Eqs. (\ref{en1}),
(\ref{fi1}) and (\ref{8.88}), we obtain:
\begin{equation}
\label{10.88}
\chi_{h}(i)=\sum_{j} \, A_{ij} {\chi_{\rm Mn}}(j)+ B_{i}
\end{equation}
where
\begin{eqnarray}
\nonumber &A_{ij}={1 \over 2} \sum_{n\sigma} \left[ \sum_{m\ne n}
f(E_{n\sigma}){{J_{m\sigma,
n\sigma}(j)\psi^{*}_{n\sigma}(i)\psi_{m\sigma}(i)+c.c.} \over
{E_{m\sigma}-E_{n\sigma}}}\right. \nonumber \\ \nonumber &
+\!\left. \beta |\psi_{n \sigma}(i)|^{2} g(E_{n\sigma}) \!\left(\!
J_{n\sigma, n\sigma}(j)\!-\!\sigma {{\sum_{m\alpha} \alpha \,
J_{m\alpha, m\alpha}(j)g(E_{m \alpha})} \over \sum_{m\alpha}g(E_{m
\alpha})} \!\right)\!\right]
\end{eqnarray}
and
\begin{eqnarray}
\nonumber B_{i}={\beta \over 4} \sum_{n\sigma} |\psi_{n\sigma}(i)|^2
g(E_{n\sigma}) \left[ \sigma {{\sum_{m\alpha} \alpha \, g(E_{m
\alpha})} \over \sum_{m\alpha}g(E_{m \alpha})} - 1 \right].
\end{eqnarray}
Here it is worth mentioning that in Eq. (\ref{10.88}), the dominant
term is $\sum_{j} A_{ij} {\chi_{\rm Mn}}(j)$ which is coming from the
indirect effect of the external field on the Mn spins.

The set of Eqs.  (\ref{3.88}) and (\ref{10.88}) relate $\chi_{\rm
  Mn}(i)$ and $\chi_{h}(i)$ to one another.  The equations for
  determining ${\chi_{\rm Mn}}(i)$ at each site $i$ are then:
\begin{eqnarray}
\label{11.8}
\sum_{j} \big[ \delta_{ij} + R_{ij} \big] {\chi_{\rm Mn}}(j) =\beta(1
- P_{i}){{B'_S}} (\beta H_{i})
\end{eqnarray}
where $ R_{ij}= \beta {B'_S} (\beta H_{i}) \sum_{k}^{}J_{ik}A_{kj}$
and $P_i =\sum_{j}^{}J_{ij} B_j$. Once we know $\chi_{\rm Mn}(i)$, the
values of $\chi_{h}(i)$ can be obtained from (\ref{10.88}). The total
susceptibilities per unit volume are then $\chi_{\rm Mn}=n_{\rm
Mn}(g\mu_{B})^{2}{\sum_{i}\chi_{\rm Mn}(i) / N_{d}}$ and
$\chi_{h}=n_{\rm Mn}(g\mu_{B})^{2}{\sum_{i}\chi_{h}(i)/ N_{d}}$.


\begin{thebibliography}{99}

\bibitem{Ohnorev} H. Ohno, J. Magn. Magn. Mat. {\bf 200}, 110 (1999).

\bibitem{Edmonds}K. W. Edmonds, P. Bogusawski, K. Y. Wang,
  R. P. Campion, S. N. Novikov, N. R. S. Farley, B. L. Gallagher,
  C. T. Foxon, M. Sawicki, T. Dietl, M. Buongiorno Nardelli and
  J. Bernholc, Phys. Rev. Lett.{\bf 92}, 037201 (2004).

\bibitem{Nazmul} A. M. Nazmul, S. Sugahara, and M. Tanaka, Phys.
Rev. B {\bf{67}}, 241308(R) (2003).


\bibitem{Beschoten} B. Beschoten, P.A. Crowell, I. Malajovich,
D.D. Awschalom, F.  Matsukura, A. Shen, and H. Ohno,
Phys. Rev. Lett. {\bf 83}, 3073 (1999).


\bibitem{Yu} K. M. Yu, W. Walukiewicz, T. Wojtowicz, I. Kuryliszyn,
X. Liu, Y. Sasaki and J. K. Furdyna, Phys. Rev. {\bf B 65}, 201303(R)
(2002).

\bibitem{Janko} G. Zar\'and and B. Janko, Phys. Rev. Lett. {\bf 89},
047201 (2002).

\bibitem{Gerg}G. A. Fiete, G. Zar\'and and K. Damle,
 Phys. Rev. Lett. {\bf 91}, 097202 (2003).

\bibitem{Brey}L. Brey and G. Gomez-Santos, Phys. Rev. B {\bf 68},
115206 (2003).

\bibitem{Zhou}C. Zhou, M.P. Kennett, X. Wan, M. Berciu and R.N. Bhatt,
  Phys. Rev. B {\bf 69}, 144419 (2004).

\bibitem{MB1} M. Berciu and R. N. Bhatt, Phys. Rev. Lett. {\bf 87},
  107203 (2000).

\bibitem{HF} M. Berciu and R. N. Bhatt, Phys. Rev. B {\bf 69}, 045202
  (2004).

\bibitem{MPK2} M. P. Kennett, M. Berciu and R. N. Bhatt,
Phys. Rev. {\bf B 66}, 045207 (2002).

\bibitem{RPA} M. Berciu and R. N. Bhatt, Phys. Rev. B {\bf 66}, 085207
  (2002).

\bibitem{MB2} M. Berciu and R. N. Bhatt, Phys. Rev. Lett. {\bf 90},
029702 (2003).

\bibitem{MPK1} M. P. Kennett, Mona Berciu and R. N. Bhatt,
  Phys. Rev. B {\bf 65}, 115308 (2002).

\bibitem{Heb} N. Theodoropoulos, A. F. Hebard, M. E. Overberg,
C. R. Abernathy, S. J. Pearton, S. N. G. Chu and R. G. Wilson,
Appl. Phys. Lett. {\bf 78}, 3475 (2001).

\bibitem{GeSi} Y. D. Park, A. T. Hanbicki, S. C. Erwin,
  C. S. Hellberg, J. M. Sullivan, J. E. Mattson, T. F. Ambrose,
  A. Wilson, G. Spanos and B. T. Jonker, Science {\bf 295}, 651
  (2002).

\bibitem{Konig} J. K\"onig, H.-H. Lin and A. H. MacDonald,
  Phys. Rev. Lett. {\bf 84}, 5628 (2000); J. Schliemann, J. K\"onig,
  H.-H. Lin and A. H. MacDonald, Appl. Phys. Lett. {\bf 78}, 1550
  (2001).

\bibitem{Amir} A. Chattopadhyay, S. Das Sarma and A. J. Millis,
  Phys. Rev. Lett. {\bf 87}, 222702 (2001).

\bibitem{Dietl} T. Dietl, H. Ohno and T. Matsukura, Phys. Rev. B {\bf
  63}, 195205 (2001).

\bibitem{Timmetal} C. Timm, F. von Oppen, and F. H\"ofling,
  Phys. Rev. B {\bf 69}, 115202 (2004); Y. Qi and S. Zhang,
  Phys. Rev. B {\bf 67}, 052407 (2003).

\bibitem{Bhatt1} R. N. Bhatt, Phys. Rev. B {\bf 24}, 3630 (1981);
  Phys. Rev. B {\bf 26}, 1082 (1982).

\bibitem{BG} A.K. Bhattacharjee and C.B. \'a la Guillaume, Solid State
  Commun.{\bf 113}, 17 (2000).


\bibitem{Timm} C. Timm, F. Sch\"afer, and F. von Oppen,
  Phys. Rev. Lett. {\bf 89}, 137201 (2002).

\bibitem{note1} The most general variational form would allow spin
flips $\sum_{i,j,\sigma}^{}h^B_{i \alpha, j\beta}
c^{\dagger}_{i\alpha} c_{j\beta}$ and non-collinearity $\sum_{i}
\vec{H}^B_{i} \cdot \vec{S}_i$ in $\hat {\cal K}$. However, a previous
study of this model \cite{HF} showed that the self-consistent
mean-field ground state is always collinear, and thus the variational
guess of Eq. (\ref{12}) is appropriate.


\bibitem{Ripka} J.-P. Blaizot and G. Ripka, {\em Quantum theory of
finite systems} (MIT Press, Cambridge, Mass., 1986).

\bibitem{Furdyna} J. Furdyna, J. Appl. Phys. {\bf 64}, R29 (1988);
  S. Lee, M. Dobrowolska, J. K. Furdyna and L. R. Ram-Mohan,
  Phys. Rev. B {\bf 61}, 2120 (2000);

\bibitem{Zem} M. A. Zudov, J. Kono, Y. H. Matsuda, T. Ikaida,
    N. Miura, H. Munekata, G. D. Sanders, Y. Sun and C. J. Stanton,
    Phys. Rev. B {\bf 66}, 161307(R) (2002).

\bibitem{MandB} M. Berciu and B. Janko, Phys. Rev. Lett. {\bf 90},
  246804 (2003).

\bibitem{MandB2} M. Berciu, T. Rappoport and B. Janko (unpublished).

\bibitem{Sch} J. Schliemann, J. K\"onig and A.H. MacDonald,
  Phys. Rev. B {\bf 64}, 165201 (2001).

\bibitem{dasS} A. Kaminski and S. Das Sarma, Phys. Rev. Lett. {\bf
 88}, 247202 (2002); Phys. Rev. B {\bf 68}, 235210 (2003).

\bibitem{Dag} M. Mayr, G. Alvarez and E. Dagotto, Phys. Rev. B {\bf
65}, 241202 (2002).


\bibitem{Adel} Adel Kassaian, {\em ``Magnetic susceptibility of
  diluted magnetic semiconductors''}, M.Sc. Thesis, (University of
  British Columbia, 2004).

\bibitem{Mayr} G. Alvarez and E. Dagotto, Phys. Rev. B {\bf 68},
 045202 (2003).

\bibitem{dasS3} D. J. Priour, Jr., E. H. Hwang, and S. Das Sarma,
Phys. Rev. Lett. {\bf 92}, 117201 (2004); S. Das Sarma, E. H. Hwang,
and A. Kaminski, Phys. Rev. B {\bf 67}, 155201 (2003).

\end{thebibliography}
\end{document}